\documentstyle[eqsecnum,aps,pre,epsf]{revtex}
\begin{document}
\draft
\title{Exact solution of diffusion limited aggregation in a 
 narrow cylindrical geometry}
\author{Boaz Kol and Amnon Aharony}
\address{Raymond and Beverly Sackler Faculty of Exact Sciences, 
	School of Physics and Astronomy,\\ Tel Aviv University, 
	69978 Ramat Aviv, Israel}
\date{\today}
\maketitle
\begin{abstract}
The diffusion limited aggregation model (DLA) 
and the more general dielectric breakdown model (DBM) are solved exactly
in a two dimensional
cylindrical geometry with periodic boundary conditions of width 2.
Our approach follows the exact evolution of the growing interface,
using the evolution matrix ${\bf E}$, 
which is a temporal transfer matrix. The 
eigenvector of this matrix with an eigenvalue of one represents the system's 
steady state.  
This yields an estimate of the fractal dimension for DLA, 
which is in good  agreement with simulations. 
The same technique is used to calculate the
fractal dimension for various values of $\eta$ in the more general 
DBM model. Our exact results are very close to the approximate results 
found by the fixed scale transformation approach.
\end{abstract}
\section{Introduction}
The problem of diffusion limited aggregation (DLA)\cite{DLA83} 
has been a subject of extensive research for the past decade and a half. 
This model produces  highly ramified and non smooth patterns which 
seem to be fractal \cite{fractal geometry}. These patterns 
have a great resemblance to those which are formed in many natural 
growth phenomena, such as viscous fingering \cite{Feder}, 
dielectric breakdown \cite{DBM84a} and many more.
A good understanding of the DLA model should help us to
explain the essential physics of these processes.

\subsection{A short description of the model}
In DLA there is a seed cluster of particles fixed somewhere; 
a particle is then released at a distance from it. 
This particle diffuses until it attempts to penetrate
the fixed cluster, in which case it gets stuck and the next particle is 
released. In this way the cluster grows. Simulations have shown that DLA 
clusters form fractal branches.
It has been shown that DLA is equivalent to the dielectric breakdown
model (DBM) with $\eta=1$ \cite{DBM84a,DBM84b}. 
This paper analyzes the DBM model. 
DBM is a cellular automaton which is defined on a lattice. It consists of the
following steps: one starts with a seed cluster of connected sites
and with a boundary surface far away from it. 
A field
$\Phi$, which corresponds to the electrostatic potential, is found 
by solving the discrete Laplace equation on a lattice,
\begin{equation}
\nabla^{2} \Phi = 0.
\label{Laplace}
\end{equation}
It is believed that the Laplace equation plays a crucial role in producing 
fractals in many physical cases, because it has no length scale and because
of its long range screening qualities. 
These growth processes are called Laplacian 
\cite{Pietronero88a,Pietronero88b,Pietronero89,Erzan95a}.
In order to solve this equation the boundary conditions must be specified. 
The aggregate is considered to have a constant potential which is 
usually set to zero. The potential gradient on the distant boundary 
is set to one in some arbitrary units. 
After solving the discrete Laplace equation (\ref{Laplace}) 
we use the field $\Phi$ to 
determine the manner in which the cluster continues to grow. The growth process
is stochastic and the growth probabilities per perimeter bond are 
determined by the local values of the electric field, equal to the 
potential difference across each bond, i.e. to 
the potential value at the sites lying across the perimeter 
bonds:
\begin{equation}
P_{b}={|\Phi_{b}|^{\eta} \over \sum_{b}|\Phi_{b}|^{\eta}}.
\label{growth probabilities}
\end{equation}
Here, $b$ is the bond index, and $\eta$ is a parameter. 
One of the perimeter bonds is chosen randomly 
according to the distribution in Eq. (\ref{growth probabilities}) and
the site across it is occupied. 
The growth continues by re-solving the Laplace equation 
(\ref{Laplace}), etc. Notice that the boundary conditions have changed a
bit because the potential on the newly occupied site is set to zero this 
time. This growth model manufactures fractal clusters without the need to fine
tune any parameter and thus differs from ordinary critical phenomena and
belongs to the class of self organized criticality (SOC) \cite{Bak87}. 
DBM can be grown in different geometries.
By geometry we refer to the dimensionality of the lattice,
to the shapes of the boundaries and to the details of the seed for growth 
(usually a point or a line for two dimensional growth). 
For instance, the case is which the distant boundary is 
a sphere is called radial boundary conditions, and
the case in which the boundary is a  distant plane at the top, while the 
seed cluster is a parallel plane at the bottom with periodic boundary 
conditions on the sides, is called cylindrical boundary conditions.

There has been considerable work on simulating DLA and measuring 
it's fractal dimension. The accepted value for the fractal dimension is 
$D=1.715$ \cite{Tolman&Meakin89} for circular DLA in two dimensions (2D)
and $D=1.66$ for infinitely wide cylindrical DLA \cite{Evertsz90}.
More details and references on numerical analysis  could be found elsewhere
\cite{Erzan95a}. A summary of values of the fractal dimension, obtained by
simulations and by theoretical approaches discussed in this paper, appears
in Table \ref{results comparison}.

\subsection{The Fixed Scale Transformation approach to DBM}
\begin{figure}
\epsfysize 6cm
\epsfbox{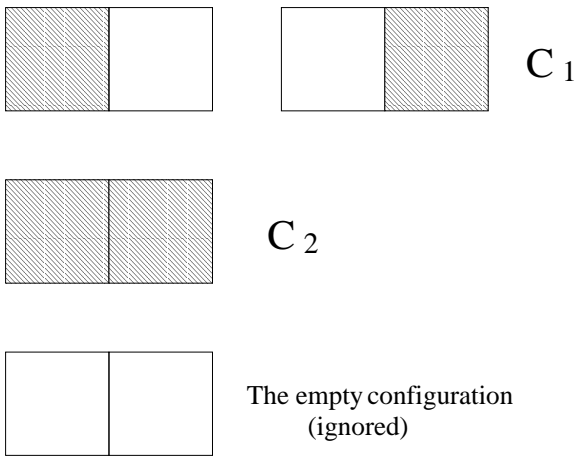}
\vspace{5mm}
\caption{Possible occupations of two adjacent sites on an intersection of 
a DBM structure which is perpendicular to the growth direction. 
These configurations have probabilities $C_1$ and $C_2$ as shown.}
\label{configurations}
\end{figure}

\begin{figure}
\epsfysize 10cm
\epsfbox{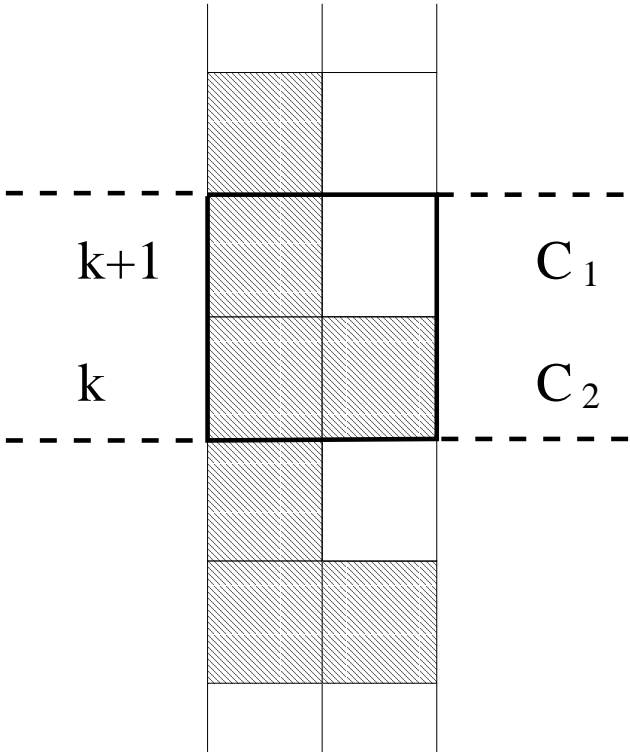}
\vspace{5mm}
\caption{The conditional probability of having a configuration $C_j$ above a 
configuration $C_i$ is the FST matrix element $M_{i,j}$. This figure shows
a $C_2$ configuration at the $k$'th row. The probability for having
a $C_1$ configuration right above it is $M_{2,1}$.}
\label{Mij}
\end{figure}

A novel approach to DBM, called the fixed scale transformation 
(FST), was introduced by Pietronero {\em et al.}
with considerable success 
\cite{Pietronero88a,Pietronero88b,Pietronero89,Erzan95a}.
Because our work 
was motivated and inspired by FST, we include a short description of
this approach, which is close in spirit to the real space renormalization group
(RSRG), but yet very different. While the RSRG 
transformation changes the scale, the FST transformation keeps the same
scale while moving in the growth direction in real space. FST analyzes the 
statistics of
the frozen structure, which is far behind the growing front. This region is
called frozen because it has very
low growth probabilities due to the screening of the Laplace equation.
The FST actually analyzes a cross section perpendicular to the growth 
direction. The most simple case studied by FST is that of the 
two dimensional cylindrical geometry \cite{Pietronero88b,Erzan95a}.
In 2D the sites on the cross section are 
gathered into pairs. 
A non-empty pair can have either one or two occupied sites. 
The probabilities for these two cases are denoted by $C_1$ and $C_2$ 
respectively, see Fig. \ref{configurations}. Then we have:
\begin{equation}
C_1+C_2=1.
\end{equation}
In FST one calculates the conditional 
probabilities of having one configuration follow another
in the growth direction. These probabilities make up the FST matrix:
\begin{equation}
\pmatrix{ C_1 \cr C_2}^{(k+1)}=
\pmatrix{ M_{1,1}&M_{2,1}\cr M_{1,2} & M_{2,2} \cr}
\pmatrix{ C_1 \cr C_2}^{(k)}
\end{equation} 
The matrix element $M_{i,j}$ represents the conditional probability of having
a configuration $j$ at the $(k+1)'{\rm th}$ row, provided there is 
a configuration $i$ at the $k'{\rm th}$ row, right below 
it,  see Fig. \ref{Mij}.
The fixed point of this transformation represents the 
asymptotic limit for the probabilities, $C^{\ast}_1$ and $C^*_2$. 
In this asymptotic limit, the average number of sites in each row is 
$\langle n \rangle = C^{\ast}_1+2C^*_2=1+C^{\ast}_2$. 
For a self similar fractal
structure, one expects that a change of scale by a factor 2 will change the
average mass of a linear cut by a factor $2^{D-1}$, where $D$ is the fractal
dimensionality of the full 2D fractal. Assuming that the above procedure
represents a coarse graining of the sites into cells of width 2, Pietronero
$et~al.$ thus concluded that $\langle n \rangle =
2^{D-1}$, i.e. 
\begin{equation}
D=1+{\log(\langle n \rangle) \over \log(2)}=1+{\log(C_1^{\ast}+2C_2^{\ast}) 
\over \log(2)}
=1+{\log(1+C_2^{\ast}) \over \log(2)}.
\label{estimate dimension}
\end{equation}
To calculate the FST matrix, one must consider all possible growth processes, 
taking account of the boundary conditions. 
Pietronero $et~al.$ computed the 
probabilities using different 'schemes'.
Here we follow one scheme, referred to as 'closed':
it is periodic with a period of 2 sites, i.e. the structure
is divided into columns, two sites wide, which are all identical. 
In order to calculate the element
 $M_{2,1}$, Pietronero $et~al.$ 
set the $k'{\rm th}$ row to be a $C_2$ configuration.
Then they considered all possible growth process which resulted in a 
configuration $C_1$ at the $(k+1)'{\rm th}$ row, and added them up with the 
corresponding
statistical weights. These statistical weights were determined by multiplying 
the probabilities for the successive growths. 
A similar procedure was done for the rest of the matrix 
elements, with the resulting fractal dimension of $D=1.55$.
Further enhancements of FST were achieved by including 
empty configurations \cite{empty}
and by working with the scale invariant growth rules \cite{Cafiero93}. 
FST also works well in 3D \cite{3dFST}.

FST is not exact, because not all possibilities are taken into account. 
For example, 
in the calculation of the element $M_{1,1}$, Pietronero {\it et al.} assume 
that there is a $C_1$ configuration at the {\it k}'th row, but they do not 
consider what happens below it. 
This is equivalent to assuming that there is a $C_2$
configuration right below it, whereas in reality there might be a few
consecutive $C_1$ 
rows. In the calculation of the element $M_{2,2}$ they assume that
there is nothing above it, whereas in reality, at the time that a $C_2$ row
is formed there may be a few $C_1$ rows above it. Moreover, the evaluation
of the elements is done by summing over a finite number of growth processes,
whereas ideally, one should sum over infinite growth processes. It is also
hard to evaluate the error in the various quantities, e.g. the FST matrix
elements $M_{i,j}$ and the fractal dimension $D$. 

\subsection{Overview}
\begin{figure}
\epsfysize 9cm
\epsfbox{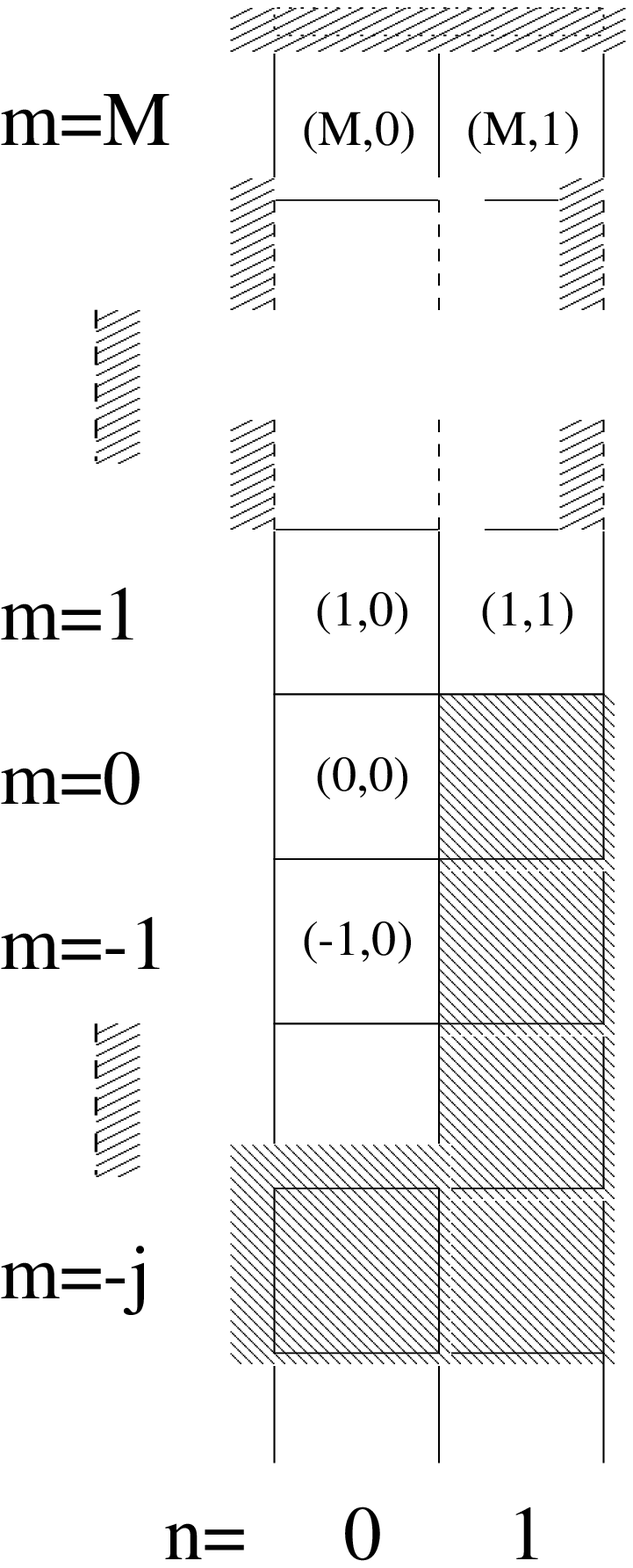}
\vspace{5mm}
\caption{The coordinates $(m,n)$ describe the location on a lattice
which is 2 sites wide. The grey sites belong to the interface of the 
aggregate which is shaped as a step whose size is $j$.}
\label{step}
\end{figure}

In this paper we solve the DBM in the geometry referred to by Pietronero 
$et~al.$ as ``closed'', i.e. in a 
2D column which is very tall but
only two sites wide, with periodic lateral boundary conditions.
Each non empty level can be either a $C_1$ or a $C_2$ configuration. 
Our solution gives the exact probabilities for $C_1$ and $C_2$, 
but not through the FST approach. 
In spite of this, we get very similar results, which
validate those of Pietronero $et~al.$
The differences between our results and those obtained with FST are 
summarized in Table \ref{results comparison} for the case $\eta=1$. 
Our approach is different from FST, because we use the interface
rather than single rows in the frozen area.
We focus our attention on the interface because it determines the solution
of the Laplace equation (\ref{Laplace}).
The solution to the equation is totally unaffected by
what happens behind the interface, i.e. by the rest of the structure. The
solution also does not depend on the history of the growth which led to the
specific interface shape.
We consider all the
possible shapes that the interface can assume, and for each possible shape
we solve the Laplace equation. 
In the case of periodic boundary conditions with period two the 
characterization
is simple: A single parameter characterizes all the possible shapes that the
interface can have.
This parameter, which we denote by $i$ or $j$, is the height
difference between the two columns, which we will call 'the step size'. 
This parameter is explained in Fig. \ref{step}. In the situation where the
two columns are of the same height it is obvious that the growth probabilities
are equal for both sides. Therefore we can assume that the growth will always
be on the same side in such an event, for instance on the right side.
This means
that the step size can always be considered as non negative.

We start by solving the Laplace equation (\ref{Laplace}) 
(the electrostatic problem) for each possible interface 
(Sec. \ref{solution}). 
First we present a general derivation in 2D with periodic
lateral boundary conditions (with a general width), then we solve for $\Phi$ 
in our special geometry of width 2 (the 'closed' scheme). We do it
by dividing the plane into two parts: the upper part, which is empty, and the
lower part, which contains the structure. We match up the two solutions by 
writing down the explicit equation for the site common to both parts. From
the potential we get the growth probabilities according to 
Eq. (\ref{growth probabilities}). 
In Sec. \ref{EP} we arrange them in a matrix, which we call
the evolution matrix, which functions as a temporal transfer matrix for this 
problem. This matrix is infinite, 
but the matrix elements $E_{i,j}$ decay exponentially
 for large $i$.
We then calculate
the steady state which is the fixed point of the evolution matrix. In 
Sec. \ref{frozen structure} we use the evolution matrix and the steady state
in order to calculate the average density of the aggregate, and therefore also
the probability $C_2$ and the fractal dimension. We continue by analyzing
the frozen structure below the growing interface. We 
observe that the frozen structure is made of a series of elements
, which we call 'hooks', and we calculate the statistics of their 
appearance. By doing so, we fully characterize the structure. 
We carry out the same procedure for a few different values of $\eta$ 
in the more general DBM model. We summarize in Sec. \ref{summary}.

\section{The solution of the electrostatic problem}
\label{solution}
\subsection{A derivation for a cylinder of arbitrary width in 
two dimensions}
\label{general_derivation}
\subsubsection{The basis solutions and the dispersion relation}

Before solving the Laplace equation for our special geometry we present
a derivation which applies to general systems with periodic 
boundary conditions in 2D.
We look at a rectangle, $M+1$ sites high and $N$ sites wide, with lateral 
periodicity. 
The Laplace equation is satisfied by every site in this rectangle. 
This is the situation in those parts in space which are unoccupied by the
aggregate.
First, we find a set of basis functions which span the linear space of 
solutions. These basis functions obey the discrete Laplace equation and
have lateral periodicity, but do not obey the boundary conditions on
the upper and lower boundaries.
We formulate the latter boundary conditions and find the solution 
which obeys them by finding the right constants for the linear combination
of the basis functions. In this process the boundary Green function will
emerge.
   
\begin{figure}
\epsfysize 8cm
\epsfbox{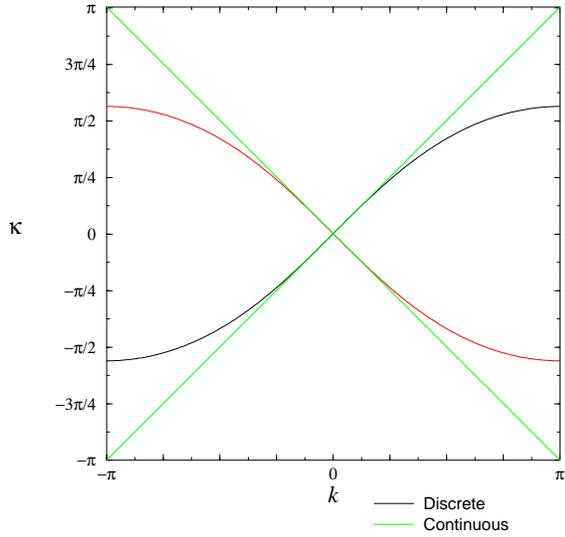}
\vspace{5mm}
\caption{The dispersion relations for the discrete and continuous Laplace
equation in 2D. The solid line shows the discrete relation 
 $\sinh(\kappa/2)=\pm \sin(k/2)$, and the dashed line shows the 
continuous relation $\kappa = \pm k$.}
\label{dispersion_fig}
\end{figure}

The discrete Laplace equation in 2D is:
\begin{eqnarray}
&&\left\{\bigl(\Phi(m,n+1)-\Phi(m,n)\bigr)-\bigl(\Phi(m,n)-\Phi(m,n-1)\bigr)
\right\}\nonumber \protect\\
&&+ \left\{\bigl(\Phi(m+1,n)-\Phi(m,n)\bigr)-\bigl(\Phi(m,n)-\Phi(m-1,n)
\bigr)
\right\}=0.
\label{discrete_rearranged}
\end{eqnarray}
Inserting an exponential solution,
\begin{equation}
\Phi(m,n)=e^{\kappa m + ikn},
\label{exponent}
\end{equation}
Eq. (\ref{discrete_rearranged})  
yields the dispersion relation
\begin{eqnarray}
\sinh^2(\kappa /2)&=&\sin^2(k/2) \\
\Rightarrow \kappa (k)=\pm 2\sinh^{-1}\left( \sin (k/2) \right)
&=&\log(q \pm \sqrt{q^2-1}),
\label{dispersion}
\end{eqnarray}
where $q \equiv 2-\cos(k)$.
This reduces to the linear dispersion relation for the continuous Laplace
equation: $\kappa=\pm k$, in the limit where the lattice constant is much 
smaller than the potential 'wave length': $\lambda \equiv 2\pi /k \gg 1$.
The relations for the discrete and continuous cases are shown in Fig. \ref
{dispersion_fig}.
The discrete case introduces an upper cutoff on the absolute value of 
$\kappa$,  
\begin{equation}
\kappa (k=\pi) \equiv \kappa_{\rm max}=2\sinh^{-1}\bigl(\sin(\pi /2)\bigr)
=\log(3+\sqrt{8})=1.7627 \dots ~.
\end{equation}
The maximum corresponds to the shortest possible wavelength, i.e. 2 sites.   
For a period $N$, the periodic boundary conditions require that $e^{ikN}=1$,
hence $k_l=2\pi l/N$ with $l=0,1,\dots ,N-1$. For each $k$ we have two 
possible $\kappa$'s: $\kappa_l \equiv \pm \kappa(k_l)=\pm \kappa(2\pi l/N)$
with $l=0,1,\dots ,N-1$. The case $k_0=0$ is special, because there is 
apparently only one solution with $\kappa_0=0$, namely 
\begin{equation}
\varphi_0(m,n)=e^{0m+i0n}=1.
\end{equation}
The second solution is obtained by considering the limit 
\begin{equation}
\psi_0(m,n)=\lim_{k, \kappa \rightarrow 0}{e^{+\kappa m+ikn}
-e^{-\kappa m+ikn} \over
2\kappa}={\partial e^{\kappa m+ikn} \over \partial \kappa}|_{\kappa,k=0}=m.
\end{equation}
The rest of the $2N-2$ basis solutions are
\begin{equation}
\varphi_l(m,n)=e^{-\kappa_lm+ik_ln}, ~~l=0,\dots,N-1 
\end{equation}
and
\begin{equation}
\psi_l(m,n)=e^{+\kappa_lm+ik_ln}, ~~l=0,\dots,N-1.
\end{equation}
\subsubsection{The solution to the boundary conditions problem and the Green
function}

The boundary conditions at the top row are that the gradient (difference) 
is uniform and equal to 1 in some arbitrary units:
\begin{equation}
\Phi(M,n)-\Phi(M-1,n)=1, ~~n=0,\dots,N-1.
\end{equation}
This condition corresponds to a uniform flux of incoming particles 
\cite{DLA83}. At the bottom the potential is 
\begin{equation}
\Phi(0,n)=f(n), ~~n=0,\dots,N-1,
\end{equation}
Where $f(n)$ is an arbitrary function.
We define
\begin{equation}
\delta \Phi(m,n) \equiv \Phi(m,n)-m.
\end{equation}
$\delta \Phi(m,n)$ also solves the discrete Laplace equation, but it obeys
different boundary conditions. At the top it has zero gradient, and at the
bottom it is the same as $\Phi(m,n)$. A set of $N$ linearly independent 
functions that obey the boundary conditions at the top 
and the discrete Laplace equation are, 
\begin{equation}
\tilde \varphi_l={\cosh(\kappa_l(M-1/2-m)) \over 
\cosh(\kappa_l(M-1/2))}e^{ik_ln},~~l=0,\dots,N-1.
\end{equation}
We now take the limit $M \rightarrow \infty$, and observe that
$\tilde \varphi_l \rightarrow \varphi_l$
(Pietronero $et~al.$ used $M=2$ in their calculations in Ref. 
\cite{Pietronero88b}). 
We have thus discarded $N$ of our basis solutions, which we denoted
by $\psi_l(m,n), ~~l=0,\dots,N-1$.
The remaining $N$ basis functions obey an orthogonality condition at the 
bottom boundary:
\begin{equation}
\sum_{n'=0}^{N-1}\varphi_l^*(0,n')\varphi_{l'}(0,n')=N\delta_{l,l'}.
\label{orthonormality}
\end{equation}
The solution will be a 
linear combination of these basis solutions:
\begin{equation}
\delta \Phi(m,n)=\sum_{l=0}^{N-1}x_l\varphi_l(m,n),
\label{linear combination}
\end{equation}
where $x_l, ~~l=0,\dots,N-1$ are $N$ complex scalars. The orthogonality 
condition (\ref{orthonormality}) implies that
\begin{equation}
Nx_{l_0}=\sum_{n'=0}^{N-1}\varphi_{l_0}^*(0,n')\delta \Phi(0,n'),
\end{equation}
and therefore
\begin{equation}
\delta \Phi(m,n)={1 \over N}\sum_{l,n'=0}^{N-1}\varphi_l^*(0,n')f(n')
\varphi_l(m,n)=\sum_{n'=0}^{N-1}f(n')G_N(n';m,n),
\end{equation}
where we introduce the boundary Green function:
\begin{equation}
G_N(n';m,n)={1 \over N}\sum_{l=0}^{N-1}\varphi_l^*(0,n')\varphi_l(m,n)
={1 \over N}\sum_{l=0}^{N-1}e^{ik_l(n-n')}e^{-\kappa_lm}.
\end{equation}
Being a linear combination of basis functions, $G_N(n';m,n)$ also obeys the
discrete Laplace equation. When $m \rightarrow \infty$ the function has zero
gradient, and at the bottom boundary it obeys
\begin{equation}
G_N(n';0,n)=\delta_{n',n}.
\end{equation}
The fact that the specified boundary conditions are real and symmetric with
respect to $n=n'$ also means that $G_N(n';m,n)$ is real and symmetric, i.e.
\begin{equation}
G_N(n';m,n)={1 \over N}\sum_{l=0}^{N-1}e^{-\kappa_lm}\cos(k_l(n-n')).
\end{equation}
The growth probabilities will be determined by the potential values near 
the interface, so only the rows $m=0$ and $m=1$ will be of importance to us.
In this formulation, the row $m=0$ is known, so we are really only interested
in the row $m=1$, which will be determined by $G_N(0;1,n)$. We therefore 
denote
\begin{equation}
g_N(n) \equiv G_N(0;1,n)={1 \over N} \sum_{l=0}^{N-1}e^{-\kappa_l}\cos(k_ln)
, ~~n=0,\dots,N-1.
\label{g_N(n)}
\end{equation}
Before proceeding we note that
\begin{equation}
\sum_{n=0}^{N-1}g_N(n)=\sum_{n=0}^{N-1}\varphi_0^*(1,n)G_N(0;1,n)
=\sum_{l=0}^{N-1}{1 \over N}=1.
\label{sum g}
\end{equation}
The final expression for the solution in a cylinder of width $N$ is
\begin{equation}
\Phi(1,n)=1+\delta \Phi(1,n)=1+\sum_{n'=0}^{N-1}g_N(|n-n'|)\Phi(0,n').
\label{potential solution}
\end{equation}

\subsection{Solution of the electrostatic problem with period 2}
We now turn to solve for $\Phi$ in our geometry (Fig. \ref{step}).
We note again that all of the structure below
the interface has no effect on the solution for $\Phi$, and hence does not
change the growth probabilities. As mentioned earlier, the interface has the
shape of a step whose height is variable. The conditions for the 
derivation of Sec. \ref {general_derivation} are not fulfilled now
because the set of sites which obey the discrete Laplace equation
do not form a rectangle.
We therefore solve the problem by dissecting
the plane into two parts; the upper part with $m\geq 0$, which is empty, and 
the lower part with $m\leq 0$, which contains the aggregate.
First, we solve the Laplace equation (\ref{discrete_rearranged}) 
for the upper and the
lower parts separately, expressing them in terms of the potential at the 
connecting site, $\Phi(0,0)$, which is denoted by $y$.
Then we write the explicit Laplace equation for the site $(0,0)$ to patch
the two parts together.
\subsubsection{The upper part solution}
The upper part $m\geq 0$ is rectangular with lateral periodicity with
$N=2$ and with gradient one for $m \rightarrow \infty$, so
we apply the general 
derivation of Sec. \ref{general_derivation}.
We have $k_l=0,\pi$ and $\kappa_l=0,\kappa_{\rm max}$ for 
$l=0,1$ respectively.
We calculate the values of the Green function using
Eq. (\ref{g_N(n)}) and Eq. (\ref{sum g}):
\begin{eqnarray}
g_2(0)&=&{1 +e^{-\kappa_{\max}} \over 2}={1 +3 -\sqrt{8} \over 2}=2-\sqrt{2} \\
g_2(1)&=&1-g_2(0)=\sqrt{2}-1.
\end{eqnarray}
The conditions at the lower boundary are $\Phi(0,n)=y,0$ for $n=0,1$
respectively, where $y \equiv \Phi(0,0)$ is yet to be determined.
We obtain the solution for the upper part by using 
Eq. (\ref{potential solution}):
\begin{eqnarray}
\Phi_{\rm up}(1,0)&=&1+yg_2(0)=1+(2-\sqrt{2})y \\
\Phi_{\rm up}(1,1)&=&1+yg_2(1)=1+(\sqrt{2}-1)y.
\end{eqnarray}

\subsubsection{The lower part solution}
Here we have to solve the potential inside the 'fjord' which is one site wide
and $j$ sites deep (Fig. \ref{step}). 
Since both sides of the 'fjord' belong to the structure, with
$\Phi=0$, the equation for the potential in the lower part is:
\begin{equation}
4\Phi_{\rm low}(m,0)=\Phi_{\rm low}(m-1,0)+\Phi_{\rm low}(m+1,0).
\end{equation}
Substituting a solution of the form $\Phi_{\rm low}(m,0)=e^{\kappa_f m}$, 
we find that 
\begin{equation}
\sinh(\kappa_f /2)= \pm 1/\sqrt{2} =\pm \sin(\pi/4),
\end{equation}
with the positive solution
\begin{equation}
\kappa_f=\kappa(k=\pi/2)=2\sinh^{-1}(1/\sqrt{2})=\log(2+\sqrt{3})=1.3170 
\dots ~
\end{equation}
The solution will be a linear combination of the two solutions:
\begin{equation}
\Phi_{\rm low}(m,0)=u_1e^{-\kappa_f m}+u_2e^{\kappa_f m},
\end{equation}
where the coefficients $u_1$ and $u_2$ are determined by
the boundary conditions:
\begin{eqnarray}
\Phi(0,0)&=&y, \nonumber \\
\Phi(-j,0)&=&0,
\end{eqnarray}
and the solution is:
\begin{equation}
\Phi_{\rm low}(m,0)=y{\sinh\bigl(\kappa_f(m+j)\bigr) \over 
\sinh(\kappa_fj)}=ye^{\kappa_fm}{1-e^{-2\kappa_f(m+j)}\over1
-e^{-2\kappa_fj}},~ m=-j,\dots,0.
\end{equation}

\subsubsection{The solution for $y \equiv \Phi(0,0)$}
We have expressed the potential for all the sites as a function of $y$. 
The value for $y$ is obtained from the Laplace equation for $(0,0)$,
\begin{equation}
4y=\Phi(-1,0)+\Phi(1,0)=y{\sinh\bigl(\kappa_f(j-1)\bigr)  \over \sinh
(\kappa_fj)} +1+(2-\sqrt{2})y. 
\end{equation}
We can simplify this a bit by expanding the term
\begin{eqnarray}
&&\sinh(\kappa_f(j-1))=\cosh(\kappa_f)\sinh(\kappa_fj)-\sinh(\kappa_f)
\cosh(\kappa_fj)=2\sinh(\kappa_fj)-\sqrt{3}
\cosh(\kappa_fj) \nonumber \\
&&\Rightarrow {\sinh(\kappa_f(j-1))\over\sinh(\kappa_fj)}=
2-\sqrt{3}-\sqrt{3}{e^{-\kappa_fj}\over\sinh(\kappa_fj)}=
2-\sqrt{3}-2\sqrt{3}{e^{-2\kappa_fj}\over 1-e^{-2\kappa_fj}},
\end{eqnarray}
yielding
\begin{equation}
y(j)=\Bigl(\sqrt{2}+\sqrt{3}+2\sqrt{3}{e^{-2\kappa_fj}
\over 1-e^{-2\kappa_fj}}\Bigr)^{-1}=y(\infty){1-e^{-2\kappa_fj}
\over 1+\beta e^{-2\kappa_fj}},
\label{y}
\end{equation}
where we denote $y(\infty) \equiv \sqrt{3}-\sqrt{2}=0.3178 \dots$
and $\beta \equiv 5-\sqrt{24}=0.1010 \dots$.
The only parameter on
which the solution depends is the step size $j$. The dependence
on $j$ is not strong; already for $j=4$ the solution is almost identical
to the solution for $j=\infty$. 

We conclude this section with a summary of the solution on the external
boundary of the cluster:
\begin{eqnarray}
\Phi_j(1,1)&=&1+g_2(1)y(j)=1+g_2(1)y(\infty){1-e^{-2\kappa_fj}
\over 1+\beta e^{-2\kappa_fj}}, \\
\Phi_j(m,0)&=&y(j){\sinh(\kappa_f(m+j)) \over \sinh(\kappa_fj)}
=y(\infty)e^{\kappa_fm}{1-e^{-2\kappa_f(j+m)} \over 1+\beta e^{-2\kappa_fj}}
,~m=-j,\dots,0.
\end{eqnarray}
The subscript $j$ denotes the step size. The 
potential is specified only for sites which are nearest neighbors to the
aggregate, and thus candidates for growth.

\section{The evolution matrix and the steady state}
\label{EP}
\subsection{The Evolution matrix ${\bf E}$}
\begin{figure}
\epsfysize 9cm
\epsfbox{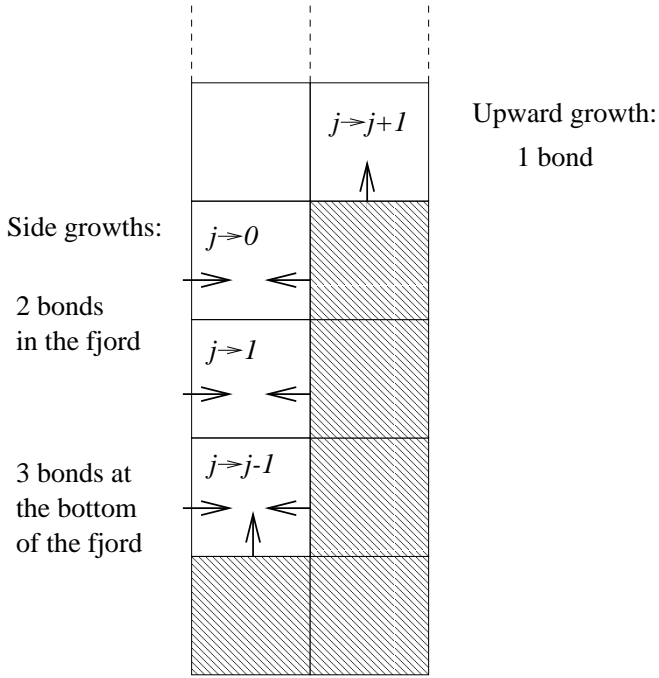}
\vspace{5mm}
\caption{Possible growth processes which change the interface from an initial
step size $j=3$ to a final size $i=4,0,1,2$. 
The growth probability is determined by the potential and the number 
of bonds associated with the site where growth is to occur.}
\label{transitions}
\end{figure}

We now proceed to calculate the growth probabilities. Growth can be 
considered as a process in which we start with a step of size $j$ and end
up with a step of size $i$, with conditional probability $E_{i,j}$ (note 
the different notation, compared to Pietronero $et~al.$'s $M_{j,i}$).
The new step size $i$ may be either equal to $j+1$ (by a growth process
in the same column), 
or smaller than $j$ (by a growth process in the adjacent column).   
The transitions are explained in Fig. \ref{transitions}. 
$E_{i,j}$ depends on the potential at the relevant site, which we denote by
$F_{i,j}$, for which we can write explicit expressions using the final results
of the previous section:
\begin{equation}
F_{i,j} \equiv \left\{ 
\begin{array}{cll}
\Phi_j(-i,0)=y(\infty)e^{-\kappa_fi}
{1-e^{-2\kappa_f(j-i)} \over 1+\beta e^{-2\kappa_fj}} &,&~0 \leq i\leq j-1 \\
\Phi_j(1,1)=1+g_2(1)y(\infty){1-e^{-2\kappa_fj} 
\over 1+\beta e^{-2\kappa_fj}} &,& ~i=j+1 \\
0&,&{\rm ~otherwise}
\end{array}
\right\} ,
~~j \geq 1.
\end{equation}
In the case $j=0$ there are two possible growth processes, in sites 
$(1,0)$ and $(1,1)$, but both of them result in a final state with $j=1$. 
The potential at these two sites is equal to one, 
hence $F_{1,0}=1$ and $F_{i,0}=0$ for $i \neq 1$. 
We note that each growth process has a 
different number of bonds associated with it: The growth upwards has one bond,
whereas all the side growths have 2 bonds (due to the periodic boundaries), 
except for the growth at the bottom site, which has 3 bonds.
This is manifested in the bond matrix element $B_{i,j}$, which is equal
to the number of bonds associated with a growth process which transforms
a step of size $j$ into a step of size $i$,
\begin{equation}
B_{i,j}=\left\{
\begin{array}{lll}
1 &,& ~i=j+1 \\
2 &,& ~0 \leq i \leq j-2 \\
3 &,& ~i=j-1 \\
0 &,& ~{\rm otherwise}
\end{array}
\right\}, ~~j \geq 1.
\end{equation}
For $j=0$ there are two bonds (leading to different sites) which 'grow' to the
state $j=1$, hence $B_{1,0}=2$ and $B_{i,0}=0$ for $i \neq 1$.
The growth probabilities are computed using Eq. (\ref{growth probabilities}):
\begin{equation}
E_{i,j}(\eta)={B_{i,j}F_{i,j}^{\eta} \over S_j(\eta)}, ~~i,j=0,\dots,\infty,
\label{etaE_(i,j)}
\end{equation} 
where we denote
\begin{equation}
S_j(\eta)=\sum_{i=0}^{\infty}B_{i,j}F_{i,j}^{\eta}, ~~j=0,\dots,\infty
\label{S_j}
\end{equation}
as the normalization factor. 

From now until Sec. \ref{eta} we only deal with the case $\eta=1$, which
corresponds to DLA. In this case, the evaluation of $S_j(\eta)$ becomes
simple, since we can use Gauss' law. 
The continuous version of the law, $\int dV \nabla^2\Phi=\oint dA 
\nabla_n\Phi$, transforms into 
\begin{equation}
\sum_{\rm bulk~sites}\nabla^2\Phi=\sum_{\rm interface~bonds}\Delta \Phi
\end{equation} 
in the discrete case. In our case the term on the left is equal to zero.
The term on the right includes contributions from the top and bottom 
boundaries only. The sum over the sides cancels because of the periodicity. 
The boundary conditions at the top require that the gradient of $\Phi$ is one,
so the sum over the top equals $N=2$. Thus, the sum over the bottom boundary
is equal $-2$, but it is also equal to minus the normalization factor, 
hence 
\begin{equation}
S_j(\eta=1)=\sum_{i=0}^{\infty}B_{i,j}F_{i,j}=2, ~~j=0,\dots,\infty.
\end{equation}
This enables us to write explicit expressions for the growth probabilities:
\begin{equation}
E_{i,j}=\left\{
\begin{array}{cll}
y(\infty)e^{-\kappa_fi}{1-e^{-2\kappa_f(j-i)}\over 1+\beta e^{-2\kappa_fj}}
&,&~0 \leq i \leq j-2 \\
{3\over 2}y(\infty)e^{-\kappa_f(j-1)}{1-e^{-2\kappa_f} \over
1+ \beta e^{-2\kappa_fj}} &,& ~i=j-1 \\
E_{\infty+1,\infty}\left(1-\alpha {e^{-2\kappa_fj} \over 1+\beta 
e^{-2\kappa_fj}} \right) &,& ~i=j+1 \\
0 &,& {\rm otherwise}
\end{array}
\right\},~~j \geq 1,
\label{AnalyticE(i,j)}
\end{equation} 
where 
\begin{equation}
E_{\infty+1,\infty}=\lim_{j\rightarrow \infty}E_{j+1,j}=
(1+g_2(1)y(\infty))/2=0.5658 \dots 
\end{equation}
and $\alpha=(1+\beta)g_2(1)y(\infty)/(2E_{\infty+1,\infty})=0.1281 \dots$. 
For $j=0$, the interface will
transform into a step of size $j=1$ with probability one, hence $E_{1,0}=1$
and $E_{i,0}=0$ for $i \neq 1$.
The values of $E_{i,j}$ are shown here for $0 \leq i,j \leq 4$, up to the 
fourth decimal digit: 
\begin{equation}
{\bf E}=\left[
\begin{array}{cccccc}
0 & 0.4393 & 0.3160 & 0.3177 & 0.3178 & \cdots \\
1 & 0 & 0.1185 & 0.0847 & 0.0851 & \\
0 & 0.5607 & 0 & 0.0318 & 0.0227 & \\
0 & 0 & 0.5655 & 0 & 0.0085 & \\
0 & 0 & 0 & 0.5658 & 0 & \\
\vdots & & & & & \ddots
\end{array}
\right].
\end{equation}

Let's examine some additional features of the matrix. 
The normalization requires that the sum of the elements in each column 
be equals to one, $\sum_{i=0}^{\infty}E_{i,j}=1$ for $j=0,\dots,\infty$.
Notice that the main diagonal is zero. 
This occurs because there is no chance
of staying with the same step size after a growth process.
The first diagonal below the main, 
which represent the probability for the step to grow larger by one, 
$E_{j+1,j}$ grows just a bit as we go down, approaching an asymptotic value of 
$E_{\infty+1,\infty}\approx .5658$ exponentially, as the third row of 
Eq. (\ref{AnalyticE(i,j)}) indicates.
The diagonal above the main represents the probabilities for growths at 
the bottom of the step, $E_{j-1,j}$, and corresponds to the second row in
Eq. (\ref{AnalyticE(i,j)}). 
These probabilities decay exponentially as the step grows deeper. 
According to the first row in Eq. (\ref{AnalyticE(i,j)}), 
the elements $E_{i,j}$ converge exponentially for
large $j$'s to a simple exponential function:
\begin{equation}
E_{i,\infty}=\lim_{j\rightarrow \infty}E_{i,j}=y(\infty)e^{-\kappa_fi}.
\label{E_(j,infty)}
\end{equation}
These probabilities relate to the transition from a very deep step to a 
step of size $i$.  

\subsection{The steady state ${\bf P^*}$}

We can describe the state of the system (the interface) using an infinitely 
long probability state vector ${\bf P}$, whose component 
$P_j, ~~(j=0,\dots,\infty)$ represents the probability of the interface to have
a step of size $j$, with
\begin{equation}
\sum_{j=0}^{\infty}P_j=1.
\label{sumP_j=1}
\end{equation}
The state with a step size $j_0$ would be described by the vector
$P_j=\delta _{j,j_0}$, e.g. the state $j_0=0$ (where the two columns are
 of equal height)  would be described by the vector ${\bf P}=(1,0,0 \dots)$.
The dynamics of the system is now described by the Master equation 
\begin{equation}
P_i(t+1)=\sum_jE_{i,j}P_j(t),
\label{master equation}
\end{equation}
or in matrix notation:
\begin{equation}
{\bf P}(t+1)={\bf EP}(t),
\label{transfer matrix}
\end{equation} 
 Equation (\ref{transfer matrix}) also shows 
that  the matrix ${\bf E}$ functions as a transfer matrix and justifies the 
name 'evolution matrix'. A state of particular interest is the steady state
which satisfies:
\begin{equation}
{\bf P^*}={\bf EP^*}.
\label{P=EP}
\end{equation}
It can be shown that if such a state exits it must be attractive,
i.e. it is reached from any initial vector ${\bf P}(0)$.
Specifically, the difference ${\bf P}(t)-{\bf P^*}$ decays 
exponentially for large $t$: 
the absolute value of all the eigenvalues of ${\bf E}$ must
be less than or equal to one. This is because ${\bf E}$ is a matrix of 
conditional
probabilities, i.e. it transforms a probability vector into a probability 
vector. If there was an eigenvalue whose absolute value was greater than one 
then after a few iterations ${\bf P}(t)$ would either contain negative 
elements or elements greater than one.
Our numerical calculations suggest that besides the eigenvalue one there is
a continuum of complex eigenvalues, all with a magnitude of 
$E_{\infty+1,\infty}$, or at least very close to it. The corresponding
eigenvectors resemble the Fourier basis (without the constant vector 
$(1,1,1,\dots)$). This can
be understood in light of the form of the matrix elements $E_{i,j}$ for large
$i$ and $j$'s; There the matrix has a Toeplitz form, because all the 
elements are practically null, besides those on the diagonal below the 
main, which have almost a constant value of $E_{\infty+1,\infty}$.
Can we be sure that a fixed point vector does exist for a general conditional
probability matrix?
From the theory of finite dimensional linear algebra it is known that 
a conditional probabilities matrix must have a fixed point,
but in the  case of an infinite number of states, a fixed point cannot be 
generally guaranteed \cite{For example}.

The calculation of the steady state is not trivial, because it requires
the manipulation of an infinite matrix. It is therefore beneficial to study
first the behavior of the steady state $P^*_j$ for large $j$'s.
From now on we will only consider the steady state and thus will omit the
superscript. The steady state equation 
(\ref{P=EP}), can be written explicitly, using (\ref{AnalyticE(i,j)}):
\begin{eqnarray}
P_i&=&E_{\infty+1,\infty}\left(1-\alpha {e^{-2\kappa_f(i-1)} \over 1+\beta 
e^{-2\kappa_f(i-1)}}\right)P_{i-1} 
+y(\infty)e^{-\kappa_fi}\sum_{j=i+1}^{\infty}
{1-e^{-2\kappa_f(j-i)} \over 1+\beta e^{-2\kappa_fj}}P_j \nonumber \\
&&+{y(\infty) \over 2}e^{-\kappa_fi}{1-e^{-2\kappa_f} \over 1+\beta 
e^{-2\kappa_f(i+1)}}P_{i+1}, ~~i \geq 2.
\label{steady state}
\end{eqnarray}
For large $i$'s the two last terms are exponentially small. If we omit the
exponential correction from the first term we find that 
\begin{equation}
P_i=E_{\infty+1,\infty}P_{i-1}+O(e^{-\kappa_fi}). 
\label{first order}
\end{equation}
The physical meaning is that very high
steps are almost always formed from a shorter step by an upward growth, 
and very seldom from higher steps by a growth deep in the 'fjord'.
We therefore use the following substitution:
\begin{equation}
P_i=x_0E_{\infty+1,\infty}^i(1+x_1e^{-\kappa_fi}+x_2e^{-2\kappa_fi}+
O(e^{-3\kappa_fi})),
\end{equation}
where the $x_i$'s are constants. Inserting this
expansion into Eq. (\ref{steady state}) one can solve for the various orders 
separately in a successive manner. For example the equation for the first 
order yields:
\begin{equation}
x_1=-{y(\infty) \over e^{\kappa_f}-1}\left(
{1 \over 1-E_{\infty+1,\infty}}-{1\over 1-E_{\infty+1,\infty}e^{-2\kappa_f}}
+{1-e^{-2\kappa_f} \over 2}E_{\infty+1,\infty}\right)=-0.1772\dots
\end{equation}
The second order equation results in 
\begin{eqnarray}
x_2 &=& {1\over e^{2\kappa_f}-1}\Biggl(\alpha e^{2\kappa_f}-y(\infty)\left(
{1 \over 1-E_{\infty+1,\infty}e^{-\kappa_f}}-{1\over 1-E_{\infty+1,\infty}
e^{-3\kappa_f}}\right) \nonumber \\
&& +{y(\infty)x_1\over 2}(e^{2\kappa_f}-1)e^{-\kappa_f}
E_{\infty+1,\infty}\Biggr)=0.1296\dots
\end{eqnarray}

In addition to this analytical expansion, it is also possible to calculate 
$P_j$ numerically.
An efficient way is to self consistently include the asymptotic 
behavior of $P_j$ and $E_{i,j}$ for $j>l$, where $l$ is an arbitrary order of 
truncation. For example, in the first
order approximation 
\begin{equation}
P_j^{(1)}=x_0E_{\infty+1,\infty}^j, ~~ j>l.
\end{equation}
We can now write a set of $(l+1)$ equations,
\begin{equation}
P_i=\sum_{j=0}^lE_{i,j}P_j+\sum_{j=l+1}^{\infty}E_{i,j}P_{j},~~i=0,\dots,l,
\label{numeric steady}
\end{equation}
in which, $P_i$ for $i=0,\dots\,l$ are $l+1$ unknown and $P_j$ for $j>l$
are approximated by $P_j^{(1)}$. Exact values of $E_{i,j}$ are used for
$0\leq i,j \leq l$, and a first order approximation is used for the rest
of the elements, i.e. 
\begin{equation}
E_{i,j}^{(1)}=\left\{
\begin{array}{ccl}
y(\infty)e^{-\kappa_fi} &,& ~l \geq i \geq 0,~j>l,~i \neq j-1 \\
{3\over 2}(1-e^{-2\kappa_f})y(\infty)e^{-\kappa_fi}\approx 1.3923
y(\infty)e^{-\kappa_fi} &,& ~i=l,~j=l+1.
\end{array}
\right.
\end{equation}
Thus we substitute
\begin{equation}
\sum_{j=l+1}^{\infty}E_{i,j}P_j\approx \left\{
\begin{array}{cl}
y(\infty)e^{-\kappa_fi}x_0E_{\infty+1,\infty}^{l+1}(1-E_{\infty+1,\infty})
^{-1} & ~0\leq i \leq l-1 \\
y(\infty)e^{-\kappa_fi}x_0E_{\infty+1,\infty}^{l+1}\left(
{1 \over 1-E_{\infty+1,\infty}}+{1-3e^{-2\kappa_f}\over 2}\right) & ~i=l
\end{array}
\right.
\end{equation}
into Eqs. (\ref{numeric steady}). We add the normalization conditions,
which now has the form
\begin{equation}
\sum_{j=0}^lP_j+{x_0E_{\infty+1,\infty}^{l+1} \over 1-
E_{\infty+1,\infty}}=1,
\end{equation}
and obtain a set of $l+2$ linear equation with $l+2$ variables ($P_j$ for
$j=0,\dots,l$ and $x_0$). The accuracy of this solution is better than 
$10^{-4}$ for $l\geq 5$. If we use the third order approximation,
$P_j^{(3)}\approx x_0E_{\infty+1,\infty}^j(1+x_1e^{-\kappa_fj}+x_2
e^{-2\kappa_fj})$, this accuracy is achieved already for $l=0$. This means
that we just have to solve two equations for $P_0$ and $x_0$ ($x_1$ and $x_2$
are explicit constants) and that the approximation $P_j^{(3)}$ is very
accurate for $j\geq 1$. Better accuracy will be achieved for a higher order
of truncation $l$ and for higher orders of asymptotic approximation for
$P_j$. Define
\begin{equation}
e_i(m) \equiv \max_{i \leq j<\infty}|P_j^{(m)}-P_j|,
\end{equation}
where $P_j^{(m)}$ is the $m$'th order approximation. It can be shown that
there exists a constant $c$ such that
\begin{equation}
e_i(m) \leq cE_{\infty+1,\infty}^ie^{-m\kappa_f}, ~~i \geq 0.
\end{equation} 
Our calculations show that $c$ is on the order of $0.01$.
The first nine  terms of the solution for $l=100$ are displayed 
in Table \ref{P(j)}. The solution also yields $x_0=0.57186\dots$
We can now check and see that Eq. (\ref{first order}) 
is fulfilled:
\begin{equation}
{P_5 \over P_4}=.5662,~{P_6 \over P_5}=.5659,~{P_7 \over P_6}=.5659,
~{P_8 \over P_7}=.5658 \approx E_{\infty +1,\infty}.
\end{equation}

\section{The complete solution}
\label{frozen structure}
\subsection{An estimation of the fractal dimension for $\eta=1$}

In this section we use the knowledge of the evolution matrix ${\bf E}$ 
and of the steady state ${\bf P}$ in order to compute the average density and
the fractal dimension of the aggregate. We start by computing the
average probability for a growth to increase the step size by one, i.e.
an upward growth,
\begin{equation}
p_{\rm up}=\langle E_{j+1,j} \rangle_j=\sum_{j=0}^{\infty}E_{j+1,j}P_j
=0.6812\dots
\label{pup}
\end{equation}
(note that this is true only after the aggregate gets to the steady state).
In practice one calculates the quantity $p_{\rm up}^{(m)}$, which is the
 numerical evaluation of $p_{\rm up}$, using the approximated
$P_j^{(m)}$ and an approximation for the elements $E_{j+1,j}$ for $j>l$.
It can be shown that $|p_{\rm up}^{(m)}-p_{\rm up}| \leq e_0(m)=ce^
{-m\kappa_f}$. It is possible to obtain much more accurate estimations using
higher orders of truncation $l$.

Similar to the argument used by Turkevich and Scher \cite{Turkevich89}, we
consider a large number of growth processes $n$ in the steady state. 
During this growth the aggregate would grow higher by $h=p_{\rm up}n$. 
The total area  covered by the new growth is $hN$ ($N=2$ is the width 
of the aggregate), thus the density is
\begin{equation}
\rho={n \over hN}={n \over p_{\rm up}nN}={1 \over p_{\rm up}N}=0.7340\dots
\label{rho}
\end{equation}
Although our model does not really produce fractal structures (due to the 
narrow width of our space), we can make an estimate of the fractal dimension
in the same way Pietronero $et~al.$ estimated it in Eq.
(\ref{estimate dimension}). In order to use this equation we have to perform
a calculation of the probabilities $C_1$ and $C_2$, which is straight 
forward:
\begin{eqnarray}
\rho &=& {C_1+2C_2 \over 2}={1+C_2 \over 2} \\
\Rightarrow C_2 &=&2\rho-1=0.4680 \dots
\label{C_2=}
\end{eqnarray} 
One can compare this exact value with the value obtained using 
the FST approach: $C_2=0.46$ \cite{Pietronero88b}. 
Plugging our value for $C_2$ into
Eq. (\ref{estimate dimension}) gives the fractal dimension:
\begin{equation}
D=1.5538.
\label{D}
\end{equation}
It is possible to bound the error in the $m$'th order numerical evaluation
of the fractal dimension: $|D^{(m)}-D| \leq \tilde{c}e^{-m\kappa_f}$,
where $\tilde{c}$ is some other constant. This means that one can obtain
any desired accuracy by using higher order evaluations. 
This result can be compared with Pietronero's: $D=1.55$ 
\cite{Pietronero88b,Erzan95a} for the closed scheme. We can make a more
exact comparison with FST by extending FST to include $30$ growth processes,
instead of $2$, and by using a high ceiling $M \gg 1$, instead of $M=2$, 
as used by Pietronero $et~al.$ In this case the values $C_2=0.4683$ and 
$D=1.5541$ are obtained. One can also compare our results to 
simulation results for the 2D cylindrical DLA, which is $D=1.60-6$ 
\cite{Pietronero88b,Erzan95a}, and to the 2D circular DLA, which is $D=1.71$ 
\cite{Pietronero88b,Erzan95a}. These results are summarized in 
Table \ref{results comparison}.

\subsection{Analysis of the frozen structure}
\begin{figure}
\epsfysize 9cm
\epsfbox{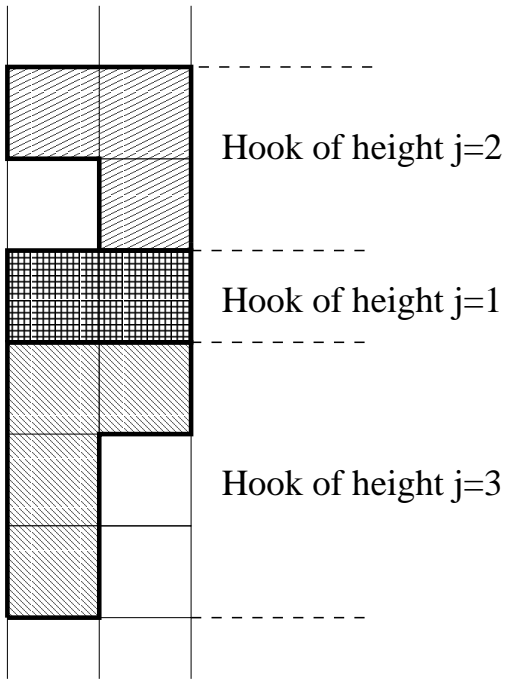}
\vspace{5mm}
\caption{The frozen structure behind the interface, composed of hook shaped
sub-structures which are laid one on top of the other.}
\label{hooks}
\end{figure}

The steady state ${\bf P}$ provides complete statistical information 
about the interface,
but it does not describe directly the properties of the structure behind the
interface, which is frozen. The key to the analysis is to understand that the 
structure is actually a series of 'hooks' of different heights, piled one on 
top of the other. 
A hook starts above a $C_2$ configuration and ends at the next $C_2$ 
configuration (including). Fig. \ref{hooks} demonstrates a few such hooks. A
full description of the structure is provided by the set of probabilities
of having a hook of height $i$, which we denote by $q(i)$. The calculation
of the $q$'s is straightforward using the steady state ${\bf P}$ and
the evolution matrix ${\bf E}$. We have to look at growth processes which 
create $C_2$ configurations (these are always side growths, which occur inside
the 'fjord'):
\begin{equation}
q(i)={1 \over 1-p_{\rm up}} \sum_{j=i}^{\infty}E_{j-i,j}P_{j}, 
~~i=1,\dots,\infty,
\label{q(i)}
\end{equation}
where $1-p_{\rm up}$ is the normalization factor because it is the average
probability for a growth to occur inside the fjord.
One can obtain an asymptotic approximation of $q(i)$ for $i \gg 1$ by
using the asymptotic approximation of $P_j$ and using a series
expansion of $E_{j-i,j}$ in terms of $e^{-2\kappa_fj}$. By doing so
one can carry out the sum in (\ref{q(i)}) and find out that
\begin{equation}
q(i)=\tilde{x}_0E_{\infty+1,\infty}^i(1+\tilde{x}_1e^{-\kappa_f i}
+\tilde{x}_2e^{-2\kappa_f i}+O(e^{-3\kappa_f i})),~~ i \geq 2,
\end{equation}
with 
\begin{eqnarray}
\tilde{x}_0&=&{x_0y(\infty) \over(1-p_{\rm up})(1-E_{\infty+1,\infty}
e^{-\kappa_f})}=0.6720, \nonumber \\
\tilde{x}_1&=&x_1{1-E_{\infty+1,\infty}e^{-\kappa_f} \over 1-E_{\infty+1,
\infty}e^{-2\kappa_f}}=-0.1567 ~\mbox{ and} \nonumber \\
\tilde{x}_2&=&(x_2-\beta){1-E_{\infty+1,\infty}e^{-\kappa_f} \over 1-
E_{\infty+1,\infty}e^{-3\kappa_f}}-1=0.9755.
\end{eqnarray} 
For $i=1$ the above expression should be multiplied by $1.5$, because 
only growths at the bottom of the fjord contribute to $q(1)$.
The eight first probabilities are presented in Table \ref{q(j)} with an 
accuracy of $10^{-4}$. They were evaluated using the sum 
(\ref{q(i)}) with very precise values of $P_j$, obtained by a high order 
truncation.
These predictions were verified using a DLA computer simulation. In
this simulation we laid 40000 hooks. Each time a hook of height $j$ was
formed a counter $\tilde q(j)$ was raised by one. Table \ref{q(j)} summarizes
the normalized results:
The fluctuations are expected to be of the order of $1/\sqrt{40000}=0.005$.
In this respect the measurement is in excellent agreement with the theory.

The $q(j)$'s give complete
information about the frozen structure, so we can also derive the fractal
dimension $D$ and  $C_2$ in terms of the $q(j)$'s:
\begin{equation}
C_2={\sum_{j=1}^{\infty}jq(j){1 \over j} \over\sum_{j=1}^{\infty}jq(j)} 
={1 \over\sum_{j=1}^{\infty}jq(j)}=0.4680 \dots
\label{C_2}
\end{equation}
Equation (\ref{C_2}) sums over the probabilities to have a row with 2 
occupied sites at the end of hooks of height $j$
(there is just one such row in a hook, the height of which is $j$).
The result in Eq. (\ref{C_2}) is the same as in Eq. (\ref{C_2=}), hence the 
estimation of the fractal dimension $D$ gives the same result as in 
Eq. (\ref{D}).

Now that we have the $q(j)$'s we can also compute the exact conditional 
probabilities for having one configuration follow another in the 
growth direction, i.e. the FST matrix elements $M_{i,j}$. 
The conditional probability for having a $C_2$ configuration above 
another $C_2$ configuration is just the probability for having a hook 
of height one. Thus, $M_{2,2}=q(1)=0.5084$. The conditional probability for
having a $C_2$ configuration above a $C_1$ configuration, $M_{1,2}$, 
can be expressed as:
\begin{eqnarray}
M_{1,2}&=&{\mbox{Probability($C_1$ at row $k$ and $C_2$ at row $k+1$)} 
\over \mbox{Probability($C_1$ at row $k$)}} \nonumber \\
&=&{\sum_{j=2}^{\infty} jq(j){1 \over j} \over 
\sum_{j=1}^{\infty} jq(j)}{1 \over C_1}={C_2 \over C_1}\left( 1-q(1) \right)
=0.4324.
\end{eqnarray}
These can be compared with $M_{2,2}=0.5056$ and $M_{1,2}=0.4142$ obtained by 
Pietronero's direct evaluation in the closed scheme 
(computed by summing up to two growths) \cite{Pietronero88b}.

Why does FST work so well? There are a few differences between our calculations
and the ones performed in Refs. \cite{Pietronero88b,Erzan95a} using FST.
First, FST uses a ceiling with $M=2$, 
instead of $M=\infty$ as is done here, but this seems to have a small effect
on the growth probabilities (less than $10^{-3}$ for $\eta=1$). In any case
one can try to implement FST also with $M=\infty$ and thus remedy this 
small effect.
Second, the fact that a relatively small number of growth processes is
considered does not change the FST matrix considerably. This effect could
also be fixed by taking into account a larger number of growth processes.
Third, in computing, for example, the conditional probabilities of having
a $C_2$ (or $C_1$) row above a given $C_2$ row, the fact that a few $C_1$ rows
may exist above the basis $C_2$ row at the time of it's formation is not taken
into account. This problem is inherent within FST and cannot be fixed in it's
framework. However, this effect is found to be small because the probability
for having two $C_1$ rows or more above a $C_2$ row at the time of it's
formation is very small (about $0.02$). Moreover, the probability of having
$j$ $C_1$ rows above a $C_2$ row at the time of it's formation decays 
exponentially as a function of $j$, with the small factor $E_{\infty+1,\infty}
e^{-\kappa_f}=0.1516$. Repeating the FST computation for the case of a 
high ceiling $M=\infty$, as in our own scheme, and accounting for as many 
as 30 growth processes, changes $D$ by $3 \times 10^{-4}$. 
This difference in $D$ is smaller by an order of magnitude from the differences
in the FST matrix elements themselves, which reflect the robustness of the FST
approximation.

\subsection{Results for different values of $\eta$}
\label{eta}
Niemeyer, Pietronero and Wiesmann introduced the DBM with the parameter $\eta$
appearing in Eq. (\ref{growth probabilities}), also referred to as the $\eta$
model \cite{DBM84a,DBM84b}. For $\eta =0$, all possible 
growths have identical probabilities, yielding a special version of the Eden 
model, which does not allow growth inside closed loops. This produces compact 
structures, i.e. the fractal and Euclidean dimensions are equal: 
$D=2$ (in our model $D(\eta=0)$ is determined by the average density and thus 
is less than $2$ because of the closed loops). For 
$\eta =1$ we get the DLA model, which has $D \approx 1.6$, and for 
$\eta =\infty$ we get a deterministic model, in which the strongest electric
field always wins, and therefore produces straight lines with $D=1$. 
We see that
as $\eta$ increases from zero to infinity, the fractal dimension $D$ decreases
from 2 to 1 continuously and monotonically. 
We can get exact results for any value of $\eta$ in the same way that we
got the exact results for $\eta=1$.
The only difference is in the values of the evolution matrix elements 
$E_{i,j}$, which are now evaluated using Eqs. (\ref{etaE_(i,j)}, \ref{S_j}). 
The steady state is then computed by solving Eqs. (\ref{P=EP}) and 
(\ref{sumP_j=1}). $p_{\rm up}$, $\rho$, $C_2$ and $D$ are evaluated using 
Eqs. (\ref{pup}, \ref{rho}, \ref{C_2=} and \ref{estimate dimension}), and
the hook height distribution $q(j)$ is found using Eq. (\ref{q(i)}). 
The solution is shown in Table \ref{eta tab}.
Note that the solution for $\eta=3$ is shown with only 3 significant digits.
This is because the higher the value of $\eta$, the slower is the convergence
of $P_j$ and $q(j)$. We used in this case a truncation scheme with $l=100$, 
and achieved an accuracy of $0.01$ for $D$.

\section{summary}
\label{summary}
We presented a complete theoretical solution of the DLA problem in a plane
with periodic boundary conditions, with a period of 2. First we identified
the possible shapes of the growing interface, as steps of varying 
heights. Then we solved the Laplace equation with the appropriate boundary
conditions. The potential defined the growth probabilities, which we inserted
into the evolution matrix ${\bf E}$. The matrix element $E_{i,j}$ was the 
conditional
probability to go from a step of size $j$ to a step of size $i$ in the next
time step, by the appropriate growth process. Next we presented the state of
the interface using an infinite vector, which we denoted by ${\bf P}(t)$.
In this notation the dynamics of the system was simply described by a
transfer matrix, see Eq. (\ref{transfer matrix}). This allowed
us to look for the steady state, which we also denoted by ${\bf P}$.
We argued that this state was attractive so that starting from any initial
condition of the system we would reach the steady state in an exponential
way. The steady state and the growth probabilities enabled us to 
calculate the average probability for upward growths, which was inversely
proportional to the average density.
We calculated the probability for having a 
filled row, $C_2$ and the complementary probability for having a half
filled row, $C_1$, and used these to obtain the fractal dimension,
$D=1.5538$. 
Our next step was to analyze the geometry of the frozen structure. We
identified it as being composed of hooks of different heights, which were 
laid one on top of the other. The frozen 
structure was fully characterized by the probabilities $q(j)$ of having a hook
of height $j$. This concluded the solution of the problem.
We also repeated the same procedure for different values of $\eta$, in the 
more general DBM model.

The solution we presented is analytical and exact, in the sense that any 
desired numerical accuracy can be achieved. The steady state vector was 
presented as a sum of exponentially decaying contributions. It was thus 
possible to bound the maximal error, with an expression that decays 
exponentially with the order of approximation.
A similar bound applied for the computed fractal dimension.
Our results are very close to those obtained by the closed scheme FST.
Our results are in excellent agreement with DLA simulations, 
which we performed in the specified geometry 
(2 sites periodic boundary conditions). The same approach
can be utilized for more complex geometries. Although it might be difficult 
to obtain exact results, our method should yield a systematic scheme of 
approximations.
\acknowledgments
We wish to thank L. Pietronero, R. Cafiero and A. Vespignani for interesting
discussions and for their cooperation. 
B.K. thanks Barak Kol for his critical review of this paper.
This work was supported by a grant from the German-Israeli Foundation (GIF).

\begin{table}
\caption{A summary of our results vs. the FST results and results obtained
from simulations}
\label{results comparison}
\begin{tabular}{l|cc}
Method&Fractal dimension $D$&Ref. \\
\tableline
Our scheme&1.5538&Present \\
\tableline
$2 \times \infty$ Simulation&1.554&-''-\\
\tableline
FST closed scheme&1.55&\cite{Pietronero88b} \\
\tableline
FST with \\ 
empty configurations \\
closed scheme&1.4655&\cite{empty} \\
\tableline
Radial simulation&1.715&\cite{Tolman&Meakin89}, \cite{Erzan95a} \\
\tableline
Cylindrical simulation&1.60-66&\cite{Evertsz90}, \cite{Erzan95a} \\
\end{tabular}
\end{table}

\begin{table}
\caption{The first 9 components of the steady state vector}
\begin{tabular}{c|ccccccccc}
$j$&0&1&2&3&4&5&6&7&8 \\
\tableline
$P_j$&.2696&.3113&.1809&.1032&.0586&.0332&.0188&.0106&.0060 \\
\end{tabular}
\label{P(j)}
\end{table}
\begin{table}
\caption{$q(j)$, the exact probability for having a hook which is $j$ sites 
tall, compared with the relative number of hooks, $\tilde q(j)$,
in a DLA computer simulation.}
\begin{tabular}{c|cccccccc}
$j$&1&2&3&4&5&6&7&8 \\
\tableline
$q(j)$&.5084&.2117&.1213&.0688&.0390&.0221&.0125&.0071 \\
\tableline
$\tilde q(j)$&.5103&.2127&.1196&.0676&.0391&.0225&.0123&.0074 \\
\end{tabular}
\label{q(j)}
\end{table}

\begin{table}
\caption{The fractal dimension for different values of $\eta$ - 
a comparison of our approach to FST (calculated up to two growth processes).
The convergence of the calculation goes like $E_{\infty+1,\infty}^l$, where
$l$ is the order of truncation.}
\begin{tabular}{c|ccccccc}
$\eta$&0&0.5&1&2&3&$\infty$ \\
\tableline
$D$&1.9144&1.7723&1.5538&1.2021&1.07&1 \\
\tableline
$D_{\rm FST}$&1.8990&1.7515&1.5418&1.1997& - &1 \\
\tableline
$E_{\infty+1,\infty}$&0&0.3128&0.5658&0.8547&0.96&1
\end{tabular}
\label{eta tab}
\end{table}

\begin{references}
\bibitem{For example}For example, the shift right (or down) operator is 
a legitimate conditional probability matrix, but it has no fixed point, 
because the number of zeros at the beginning
of the vector grows by one when applying this transformation.
\end{references}
\end{document}